\documentclass[preprint,12pt]{elsarticle}

\usepackage{amssymb}
\usepackage{chemformula} % Formula subscripts using \ch{}
\usepackage[T1]{fontenc} % Use modern font encodings
\usepackage{amsmath}
\usepackage{amssymb}
\usepackage{bibentry}
\usepackage{bm}
\usepackage{filecontents}
\usepackage{hyperref}
\usepackage{graphicx}
\usepackage{dcolumn}
\usepackage{listings}
\usepackage{xcolor} %custom colours
\usepackage{mdframed} %nice frames
\usepackage{placeins}
\usepackage[caption=false, font=large]{subfig}

\definecolor{light-gray}{gray}{0.95} %the shade of grey that stack exchange uses

\usepackage[caption=false, font=large]{subfig}
\newcounter{bla}

\newcommand{\x}[0]{\mathbf{x}}

\newcommand{\F}[0]{\mathrm{\mathbf{F}}}

\journal{Computer Physics Communications}

\begin{document}

\begin{frontmatter}



\title{GROMACS Implementation of Free Energy Calculations with Non-Pairwise Variationally Derived Intermediates}


\author[a]{Martin Reinhardt}
\author[a]{Helmut Grubm\"uller\corref{author}}

\cortext[author] {Corresponding author.\\\textit{E-mail address:} hgrubmu@gwdg.de}
\address[a]{Max Planck Institute for Biophysical Chemistry, Am Fassberg 11, 37077 G\"ottingen, Germany}
%\address[b]{Second Address}

\begin{abstract}
	
Gradients in free energies are the driving forces of physical and biochemical systems. To predict free energy differences with high accuracy, Molecular Dynamics (MD) and other methods based on atomistic Hamiltonians conduct sampling simulations in intermediate thermodynamic states that bridge the configuration space densities between two states of interest ('alchemical transformations'). For uncorrelated sampling, the recent Variationally derived Intermediates (VI) method yields optimal accuracy. The form of the VI intermediates differs fundamentally from conventional ones in that they are non-pairwise, i.e., the total force on a particle in an intermediate states cannot be split into additive contributions from the surrounding particles. In this work, we describe the implementation of VI into the widely used GROMACS MD software package (2020, version 1). Furthermore, a variant of VI is developed that avoids numerical instabilities for vanishing particles. The implementation allows the use of previous non-pairwise potential forms in the literature, which have so far not been available in GROMACS. Example cases on the calculation of solvation free energies, and accuracy assessments thereof, are provided.

%A submitted program is expected to satisfy the following criteria: it must be of benefit to other physicists, or be an exemplar of good programming practice, or illustrate new or novel programming techniques which are of importance to computational physics community; it should be implemented in a language and executable on hardware that is widely available and well documented; it should meet accepted standards for scientific programming; it should be adequately documented and, where appropriate, supplied with a separate User Manual, which together with the manuscript should make clear the structure, functionality, installation, and operation of the program.

%Your manuscript and figure sources should be submitted through Editorial Manager (EM) by using the online submission tool at \\
%https://www.editorialmanager.com/comphy/.

%In addition to the manuscript you must supply: the program source code; a README file giving the names and a brief description of the files/directory structure that make up the package and clear instructions on the installation and execution of the program; sample input and output data for at least one comprehensive test run; and, where appropriate, a user manual.

%A compressed archive program file or files, containing these items, should be uploaded at the "Attach Files" stage of the EM submission.

%For files larger than 1Gb, if difficulties are encountered during upload the author should contact the Technical Editor at cpc.mendeley@gmail.com.

\end{abstract}


\end{frontmatter}

% \linenumbers

% All CPiP articles must contain the following
% PROGRAM SUMMARY.

{\bf PROGRAM VERSION SUMMARY}
  %Delete as appropriate.

\begin{small}
\noindent
{\em Program Title: GROMACS-VI-Extension}                                          \\
{\em CPC Library link to program files:} (to be added by Technical Editor) \\
{\em Developer's respository link:} https://www.mpibpc.mpg.de/gromacs-vi-extension and https://gitlab.gwdg.de/martin.reinhardt/gromacs-vi-extension\\%(if available) \\
{\em Code Ocean capsule:} (to be added by Technical Editor)\\
{\em Licensing provisions:} LGPL \\
{\em Programming language:} C++14, CUDA                              \\
{\em Supplementary material:} 
All topologies and input parameter files required to reproduce the example cases in this work, as well as user and developer documentation will be provided online together with the source code.  \\
  % Fill in if necessary, otherwise leave out.
{\em Journal reference of previous version:}* M.J. Abraham, T. Murtola, R. Schulz, S. Pall, J.C. Smith, B. Hess, E. Lindahl, GROMACS: High performance molecular simulations through multi-level parallelism from laptops to supercomputers, \textit{SoftwareX}, 1-2 (2015)  \\
  %Only required for a New Version summary, otherwise leave out.
{\em Does the new version supersede the previous version?:} No   \\
  %Only required for a New Version summary, otherwise leave out.
{\em Reasons for the new version:*} Implementation of variationally derived intermediates for free energy calculations\\
  %Only required for a New Version summary, otherwise leave out.
{\em Summary of revisions:}*\\
  %Only required for a New Version summary, otherwise leave out.
%(approx. 50-250 words)  
{\em Nature of problem:} The free energy difference between two states of a thermodynamic system is calculated using samples generated by simulations based on atomistic Hamiltonians. Due to the high dimensionality of many applications as in, e.g., biophysics, only a small part of the configuration space can be sampled. The choice of the sampling scheme critically affects the accuracy of the final free energy estimate. The challenge is, therefore, to find the optimal sampling scheme that provides best accuracy for given computational effort.\\
  %Describe the nature of the problem here. \\
{\em Solution method(approx. 50-250 words):}
Sampling is commonly conducted in intermediate states, whose Hamiltonians are defined based on the Hamiltonians of the two states of interest. Here, sampling is conducted in the variationally derived intermediates states that, under the assumption of uncorrelated sample points, yield optimal accuracy. These intermediates differ fundamentally from the common intermediates in that they are non-pairwise, i.e., the forces on a particle are only additive in the end state, whereas the total force in the intermediate states cannot be split into additive contributions from the surrounding particles. 
  %Describe the method solution here.
{\em Additional comments including restrictions and unusual features (approx. 50-250 words):}\\
  %Provide any additional comments here.

%* Items marked with an asterisk are only required for new versions
%of programs previously published in the CPC Program Library.\\
\end{small}

\section{Introduction}
\label{sec:introduction}

Thermodynamic systems are driven by free energy gradients. Hence, knowledge thereof is key to the molecular understanding of a wide range of biophysical and chemical processes, as well as to applications in the pharmaceutical \cite{Konc2015, Christ2014, Armacost2020} and material sciences \cite{Rickman2002, Vogiatzis2020, Swinburne2018}. Consequently, \textit{in silico} calculations of free energies are popular in providing complementary insights to experiments or assisting the selection of chemical compounds in the early stages of drug discovery projects. \\

The microscopic calculation of the free energy, 

\begin{align}
\Delta G &= -\beta^{-1} \ln Z \\
         &= -\beta^{-1} \ln \int_{-\infty}^{\infty} e^{-\beta H(\x)} d\x \; ,
\end{align}

requires integration over all positions $\x$ of all particles in the system, 
where $Z$ denotes the partition sum, $\beta = 1\slash (k_B T)$ the thermodynamic $\beta$, $k_B$ the Boltzmann constant, $T$ the temperature and $H(\x)$ the Hamiltonian. As an exact integration is not feasible for high-dimensional $\x$ in case of many particles, sampling based approaches such as Monte-Carlo (MC) or Molecular Dynamics (MD) simulations are commonly used. Furthermore, in practice, it oftentimes suffices to know only the free energy difference between two states, which can be calculated much more accurately. The most basic approach, 

\begin{align}
\Delta G_{A, B} = -\beta^{-1}\ln \left\langle e^{-\beta[H_B(\x) - H_A(\x)]} \right\rangle_A
\label{eq:zwanzig}
\end{align}

rests on the Zwanzig formula \cite{Zwanzig1954}. The brackets $\langle \rangle_A$ indicate an ensemble average over $A$ is calculated. More recent methods with close relations to Eq.~(\ref{eq:zwanzig}) that use samples from both $A$ and $B$ are the Bennett Acceptance Ratio (BAR) and multistate BAR (MBAR) method \cite{Bennett1976, Shirts2008} methods. \\

For sampling based approaches, the accuracy of a free energy difference estimate between two states $A$ and $B$ generally improves when sampling is not only conducted in $A$ and $B$, but also in intermediate states. Commonly, a mostly linear interpolation between the end state Hamiltonians $H_A(\x)$ and $H_B(\x)$ is used,

\begin{align}
H_{lin}(\x, \lambda) = (1-\lambda) H_A(\x, \lambda) + \lambda H_B(\x, \lambda) \; ,
\label{eq:linear_interpolation}
\end{align} 

where $\lambda \in [0, 1]$ denotes the path variable. The $\lambda$ dependence of the end state Hamiltonians enables the use of soft-core potentials \cite{Beutler1994, Zacharias1994, Steinbrecher2007} that avoid divergences in case of vanishing particle for, e.g., the calculation of solvation free energies (where the molecules ``vanishes'' from solution). A step-wise summation,

\begin{align}
\Delta G_{AB} = \sum_{i=1}^{N-1}\Delta G_{i, i+1}
\label{eq:stepwise}
\end{align}

yields the total free energy difference, where $N$ denotes the total number of states. In the sum of Eq.~(\ref{eq:stepwise}), $i=1$ corresponds to state $A$ and $i=N$ to state $B$, respectively. Alternatively, for many steps the difference can be calculated with Thermodynamic Integration (TI) \cite{Kirkwood1935},

\begin{align}
\Delta G_{AB} = \int_{0}^{1}\left\langle \frac{\partial H(\x, \lambda)}{\partial \lambda} \right\rangle_\lambda d\lambda \;. 
\end{align}

%A different methodological class are non-equilibrium approaches. Jarzynski \cite{Jarzynski1997} derived the identity 

%\begin{align}
%e^{-\beta \Delta G_{AB}} = \left\langle e^{-\beta \Delta W} \right\rangle_A \;.
%\label{eq:jarzynski}
%\end{align}

%Here, the brackets indicate an average over all transitions starting from $A$ and ending at $B$, and $W$ the required work for these transitions. The identity requires that the starting configurations are drawn from an equilibrium ensemble of $A$. The work values are obtained similarly to TI by integration along $\lambda$, 

%\begin{align}
%W = \int_{0}^{1}\left\langle \frac{\partial H(\x, \lambda)}{\partial \lambda} \right\rangle_\lambda d\lambda \;.
%\end{align}

%A number of estimators \cite{Nanda2005, Maragakis2008, Goette2009} have been derived based on the Crooks equality \cite{Crooks1998, Crooks1999, Chelli2007} that utilize transitions going in both directions. \\

Importantly, advantageous definitions of intermediate states exist that go beyond the definition of Eq.~(\ref{eq:linear_interpolation}). For example, variationally derived intermediates (VI) \cite{Reinhardt2020, Reinhardt2020a} minimize the mean squared error (MSE) of free energy estimates using FEP and BAR. An easily parallelizable approximation for a small number of states is

\begin{align}
\resizebox{.95\hsize}{!}{$H_{VI}(\x, \lambda) = -\frac{1}{2\beta}\ln\bigg\{(1-\lambda) \exp\Big[-2\beta H_A(\x)\Big] + \lambda \exp\Big[-2\beta \Big(H_B(\x)-C\Big)\Big]\bigg\} $}\;,
\label{eq:vi_approx}
\end{align} 

where, similar to BAR, the free energy difference estimate is optimal if $C\approx \Delta G$. It is similar in shape to the minimum variance path (MVP) \cite{Gelman1998, Blondel2004, Pham2012} for TI (2 vs $1\slash2$ in the exponents). Enveloping Distribution Sampling (EDS) \cite{Christ2007, Christ2008}, and extensions such as Accelerated EDS \cite{Perthold2018, Perthold2020} use a reference potential similar in shape to Eq.~(\ref{eq:vi_approx}) to calculate the free energy difference between two or more end states. \\

\begin{figure*}[t!]
	\centering
	\includegraphics[width=0.9\textwidth]{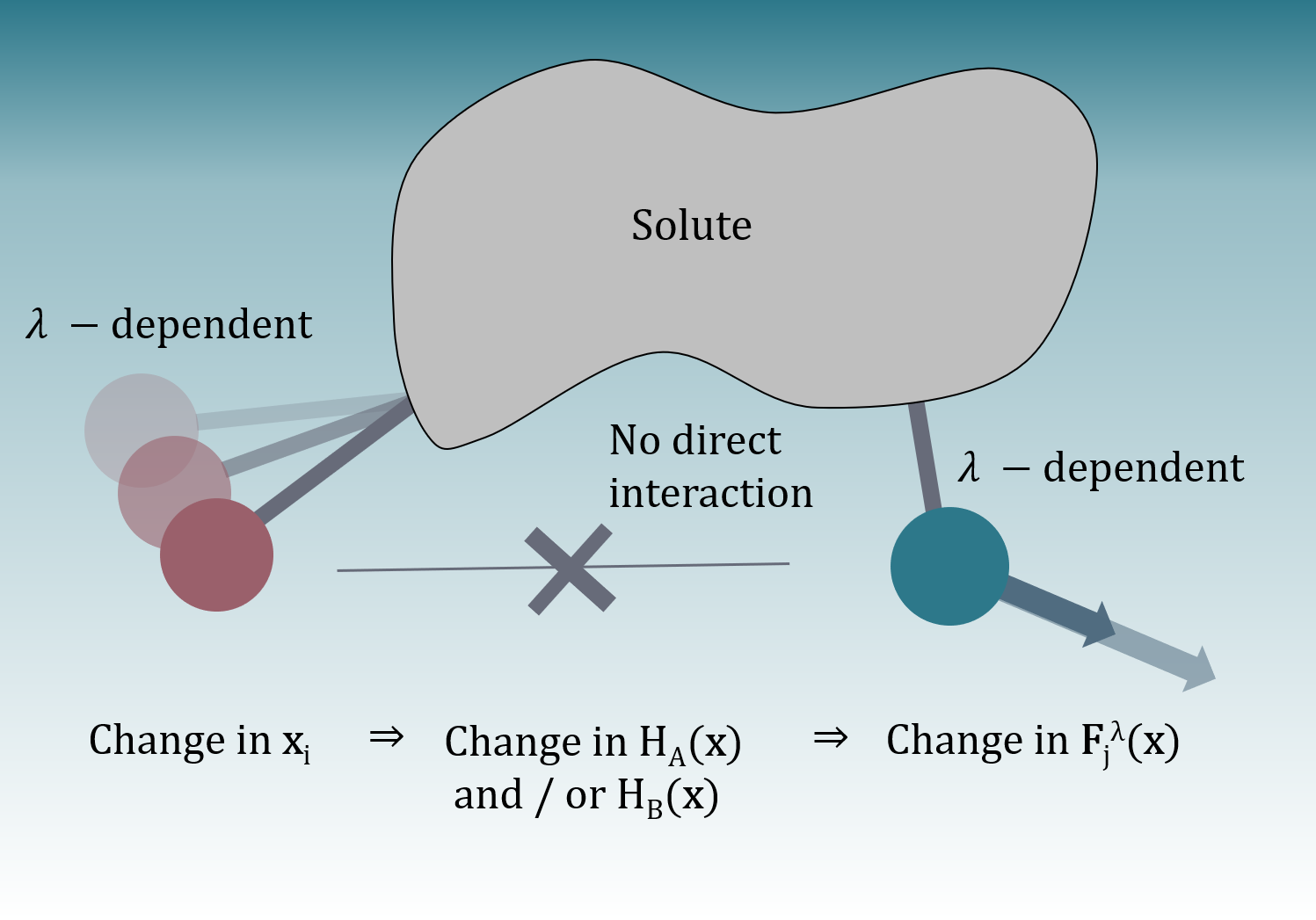}
	\caption{Non-pairwise potentials and forces in VI intermediates. Two particles $i$ and $j$ (red and blue, respectively) are considered that are $\lambda$ dependent, i.e., their interaction potential differs between $A$ and $B$. It is assumed that direct interactions between $i$ and $j$ in both $A$ and $B$ are negligible. If particle $i$ changes its position, then $H_A(\x)$ and / or $H_B(\x)$ change accordingly, and so does $H_{VI}(\x, \lambda)$. Due to the form of the VI sequence, the derivative, and therefore, the force on particle $j$ changes.}
	\label{fig:non_pairwise}
\end{figure*}

Note a particular characteristic of the VI sequence and related methods, which is illustrated in Fig.~\ref{fig:non_pairwise}: Its Hamiltonians cannot be formulated as the pair-wise sum of interaction potentials for all particles. To see this, consider the force on particle~$j$ (blue), obtained through the derivative of Eq.~(\ref{eq:vi_approx}). It still depends on the full Hamiltonians of the end states. The consequence can be understood by considering a particle $i$ (red), with $\lambda$ dependent parameters, positioned at a distance $r_{ij}$ so large such that all direct interactions between $i$ and $j$ are negligible. However, when particle $i$ changes its position with respect to its neighboring particles, the end states Hamiltonians also change, and, therefore, so does the force on particle $j$. \\

In this work, we, firstly, describe our implementation of the VI approach, and, by extension, also the MVP and basic principles of the EDS methods for two end states, into GROMACS \cite{Abraham2015, Pronk2013, VanDerSpoel2005}. It is among the most widely used MD software packages; however, none of the above approaches are available so far in GROMACS. Secondly, we introduce an approach to avoid singularities for vanishing particles with VI. 

\section{Avoiding End State Singularities}
\label{sec:theory}

%Using the terminology employed in the GROMACS documentation, we refer to these as the foreign lambda states. 

\begin{figure*}
	\centering
	\includegraphics[width=\textwidth]{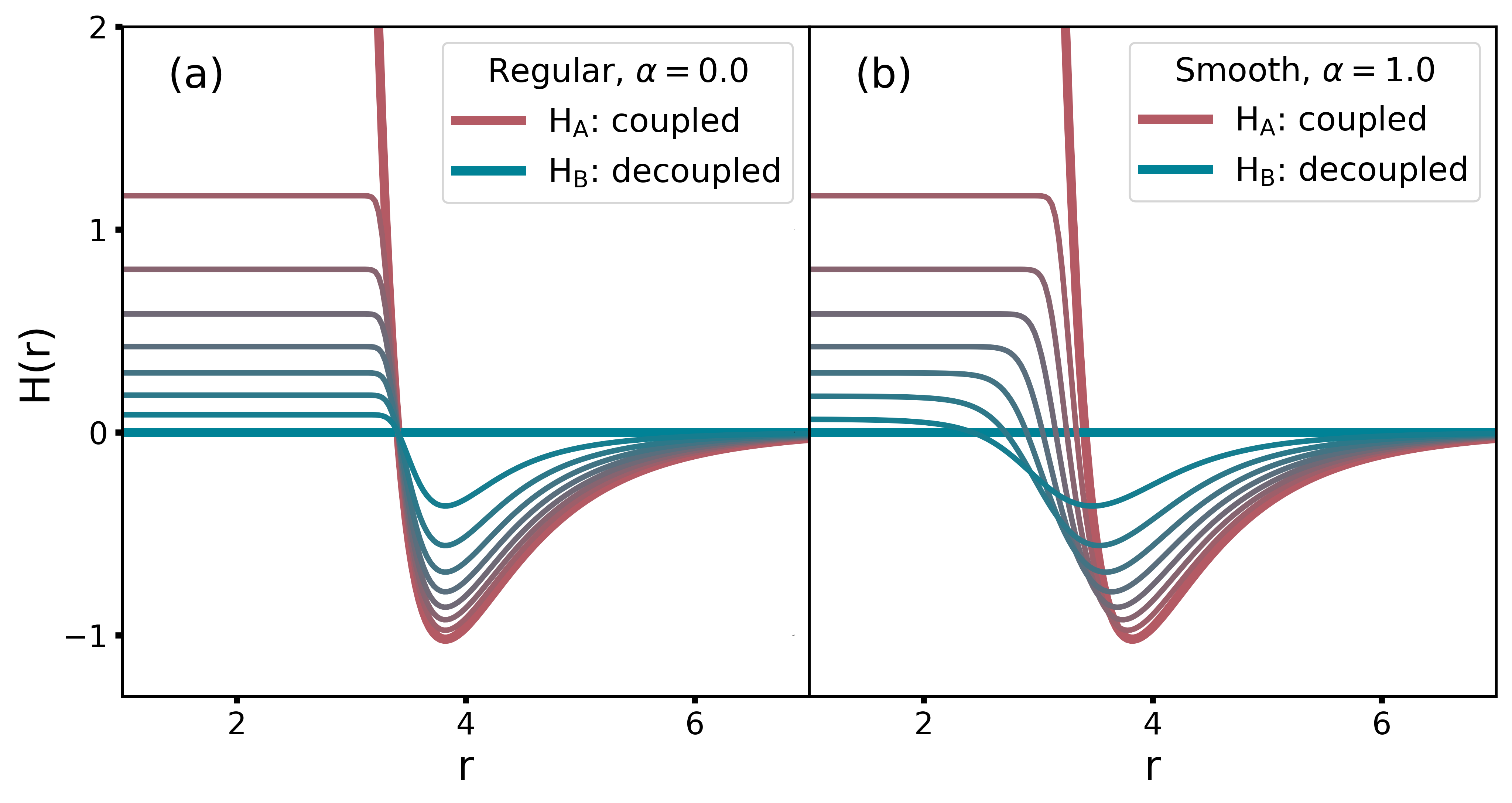}
	\caption{Intermediate VI states for a vanishing particle system. The thick red line shows the Lennard-Jones potential between two particles. The blue one shows the decoupled end state, i.e., the particles don't ``see'' each other anymore. The interpolated colors represent the intermediate states. (a) The VI sequence without and (b) with $\lambda$ dependent end states.}
	\label{fig:softcore}
\end{figure*}

%An approximated pathway to the ideal solution for finite numbers of intermediates reads as:
%\begin{align}
%H(\x, \lambda) = -\frac{1}{2\beta}\ln\bigg\{(1-\lambda) \exp\Big[-2\beta H_A(\x)\Big] + \lambda \exp\Big[-2\beta \Big(H_B(\x)-C\Big)\Big]\bigg\} .
%\label{eq:vi_approx}
%\end{align} 
Interestingly, the VI sequence, Eq.(\ref{eq:vi_approx}), already exhibits soft-core characteristics for vanishing particles, as shown in Fig.~(\ref{fig:softcore})(a) on the example of a two-particle Lennard-Jones (LJ) potential. However, divergences can still occur when configurations from the decoupled states are evaluated at foreign states, i.e., the ones that no sampling is conducted in, but that the Hamiltonian is evaluated at such as, e.g., state $B$ in Eq.~(\ref{eq:zwanzig}). Furthermore, when two particles start to overlap, very small changes in their separation $r$ lead to large changes in force, which causes instabilities due to finite integration steps.\\

 To avoid these divergences, a dependence of the end state Hamiltonians on $\lambda$ analogous to common soft-core potentials \cite{Beutler1994} is introduced, i.e., $H_A = H_A(\x, \lambda)$ with $H_A(\x, 0) = H_A(\x)$, and $H_B = H_B(\x, \lambda)$, with $H_B(\x, 1) =  H_B(\x)$, respectively. For two particle $i$ and $j$ with distance $r_{ij}$, the Coulomb and Lennard-Jones interactions in state $A$ and $B$ are calculated based on the modified distances $r_A$ and $r_B$, respectively, that are defined as
 
\begin{align}
r_A(r_{ij}, \lambda) &= \big(\alpha \sigma_{ij}^6 \lambda^p + r_{ij}^6\big)^{\frac{1}{6}} \,,\\
r_B(r_{ij}, \lambda)  &= \big(\alpha \sigma_{ij}^6 (1-\lambda)^p + r_{ij}^6\big)^{\frac{1}{6}} \,,
\end{align}

where $\alpha$ and $p$ are soft-core parameters to be specified by the user, and $\sigma_{ij}$ the Lennard-Jones parameter in the coupled state. For a system of two Lennard-Jones particles, Fig.~(\ref{fig:softcore}) shows the resulting VI states without (a) and with (b) the use of $\lambda$ dependent end states. As can be seen, the transition to the overlap region becomes markedly smoother. \\

Secondly, for increasingly complex molecules, the likelihood of barriers between the relevant parts of configuration space of the end states rises. Aside of additional techniques such as replica exchange, or meta-dynamics, the factor 2 in the exponent can be replaced by a user specific smoothing factor $s$ introduced in the EDS \cite{Christ2007, Christ2008} method. %, or, equivalently, increasing the effective temperature for exponential mixing. 
In the limit of small $s$, a series expansion of the exponential terms yields the conventional pathway, i.e., Eq.~(\ref{eq:linear_interpolation}). The modified VI sequence thus reads as

\begin{align}
\begin{split}
H_{VI}(\x, \lambda) = -\frac{1}{s\beta}\ln\bigg\{&(1-\lambda) \exp\Big[-s\beta H_A(\x, \lambda)\Big] \\
&+\lambda \exp\Big[-s\beta \Big(H_B(\x, \lambda)-C\Big)\Big]\bigg\} \;.\\
\end{split}
\label{eq:vi_all}
\end{align}

The force on particle $i$,
 
\begin{align}
\F^{VI}_i(\x, \lambda) =\,& -\frac{\partial H_{VI}(\x, \lambda)}{\partial \x_i} \\
\begin{split}
=\,& \exp\big[s\beta H_{VI}(\x, \lambda)\big] \\
& \bigg\{(1-\lambda) \exp\big[-s\beta H_A(\x, \lambda)\big]\,\F^A_i(\x) \\
&+\lambda \exp\big[-s\beta \big(H_B(\x, \lambda)-C\big)\big]\,\F^B_i(\x)\bigg\} \;,
\end{split}
\end{align}

 in the intermediate state characterized by $\lambda$, depends on both $H_A(\x, \lambda)$ and $H_B(\x, \lambda)$, as well as on the sum of the forces, $\F^A_i(\x)$ and $\F^B_i(\x)$ on particle $i$ in end state $A$ and $B$, respectively. \\

Along similar lines, the derivate 

\begin{align}
	\begin{split}
		\frac{\partial H_{VI}(\x, \lambda)}{\partial \lambda} 
		=\,& \frac{\exp\big[s\beta H_{VI}(\x, \lambda)\big]}{\beta s} \\
		& \bigg\{\left((1-\lambda) s\beta \frac{\partial H_A(\x, \lambda)}{\partial \lambda} + 1 \right)\exp\big[-s\beta H_A(\x, \lambda)\big] \\
		&+\left( \lambda s\beta \frac{\partial H_B(\x, \lambda)}{\partial \lambda} - 1 \right)\exp\big[-s\beta (H_B(\x, \lambda) - C)\big]\bigg\} \,
	\end{split}
	\label{eq:dhdl_sim}
\end{align}

depends on the derivatives ${\partial H_A(\x, \lambda) \slash \partial \lambda}$ and ${\partial H_B(\x, \lambda) \slash\partial \lambda}$ in the end states. Equation~(\ref{eq:dhdl_sim}) is used for TI. \\

Due to the dependence of Eq.~(\ref{eq:vi_all}) on $C$, where the accuracy is optimal if $C\approx \Delta G_{AB}$, the free energy difference has to be determined in an iterative process,
\begin{align}
C^{n+1} = \Delta G_{AB'} + C^n \;,
\end{align}
where $C^n$ denotes the free energy guess at iteration step $n$. The free energy difference $\Delta G_{AB'}$ is obtained from simulations between state $A$ and $B'$, where the latter denotes the end state shifted by the constant $C$, i.e., that is governed by $H_B'(\x, \lambda) = H_B(\x, \lambda) - C$. The difference $\Delta G_{AB'}$ converges to zero, such that the desired quantity $\Delta G_{AB} = \Delta G_{AB'} + C^n \approx C^n$ at the end of the iteration process. 

\section{Program Structure and Usage}
\label{sec:program}

The end states Hamiltonians,
\begin{align}
H_A(\x, \lambda) &= H_A^{\lambda}(\x, \lambda) + H^c(\x) \label{eq:ha_split}\\
H_B(\x, \lambda) &= H_B^{\lambda}(\x, \lambda) + H^c(\x) \label{eq:hb_split}, 
\end{align}

can be split into the $\lambda$-dependent energy contributions $H_A^{\lambda}(\x, \lambda)$ and $H_B^{\lambda}(\x, \lambda)$, respectively, and the common contributions summarized by $H^c(\x)$ that are equal in both end states, such as water-water interactions. To calculate $H_A(\x, \lambda)$ and $H_B(\x, \lambda)$, GROMACS only evaluates the $\lambda$-dependent contributions separately for the end states, whereas $H^c(\x)$ is calculated only once. Note that, due to the $\lambda$ dependence of the end states, $H_A^{\lambda}(\x, \lambda)$ and $H_B^{\lambda}(\x, \lambda)$ differ for different intermediates for $\alpha>0$.\\

The same holds for the VI sequence, Eq.~(\ref{eq:vi_all}). Inserting Eqs.~\ref{eq:ha_split} and \ref{eq:hb_split}, yields 
\begin{align}
H_{VI}(\x, \lambda) = H_{VI}^\lambda(\x, \lambda) + H^c(\x) \,,
\end{align}

where $H_{VI}^\lambda(\x, \lambda)$ is described by Eq.~(\ref{eq:vi_all}), where the end states Hamiltonians $H_A(\x, \lambda)$ and $H_A(\x, \lambda)$ have been replaced by the parts $H^\lambda_A(\x, \lambda)$ and $H^\lambda_B(\x, \lambda)$, respectively, that only sum over $\lambda$-dependent interactions. The same principle applies to the calculation of the forces and $\lambda$-derivatives. Therefore, the computational effort of VI is very close to the using conventional intermediates.\\

However, in the current GROMACS implementation structure, all force and energy contributions from different interaction types are interpolated between the end states right after they have been calculated, i.e., the overall calculation has the form, 

\begin{align}
H_{lin}^\lambda(\x, \lambda) \: &= \sum_{\substack{interaction \\type \;\; k  }} ... \sum_{\substack{particles \\i, j}} (1-\lambda)H^ k_A(\x_{i,j}, \lambda) + \lambda H^k_B(\x_{i,j}, \lambda) \label{eq:ham_sim}\\
F^i_\lambda(\x) \: &= \sum_{\substack{interaction \\type \;\; k  }} ... \sum_{\substack{particles \;j}} (1-\lambda)F^k_A(\x_{i,j}, \lambda) + \lambda F^k_B(\x_{i,j}, \lambda) \;.\label{eq:force_sim}
\end{align}

Whereas this has the least memory requirement, for VI, the full Hamiltonians and forces in the end states need to be known before the individual forces can be calculated. Therefore, the end states Hamiltonians and forces are stored separately. After all $\lambda$-dependent contributions have been collected, first the Hamiltonian and subsequently the forces are calculated. \\

%(xxx Add which functions it is added to?)
The implementation was built based on the GROMACS 2020 version 1 (forked on October 19th, 2019 from the master branch of the developer's repository). VI can be used with the new following entries in the mdp (i.e., input parameter) file:\\

\begin{lstlisting}[backgroundcolor = \color{light-gray}]
variational-morphing = 1 
smoothing-factor     = 2. 
deltag-estimate      = 10.3  ; in kJ / mol
\end{lstlisting}

Furthermore, the option 
\begin{lstlisting}[backgroundcolor = \color{light-gray}]
nstcalcenergy        = 1
\end{lstlisting}
should be set, as the force calculation requires the Hamiltonians of the end state. The $\lambda$ dependence of the end state Hamiltonians for VI are controlled via the already existing soft-core infrastructure,

\begin{lstlisting}[backgroundcolor = \color{light-gray}]
sc-alpha             = 0.7
sc-r-power           = 6
sc-coul              = no
sc-sigma             = 0.3
\end{lstlisting}

%Coulomb soft-core is generally not required for VI even when decoupling Coulomb and Lennard-Jones interactions simultaneously. \\

By nature of Eq.~\ref{eq:vi_all}, the transformation only takes place along a single $\lambda$ variable, to be specified by the mdp parameter \texttt{fep-lambdas}. As such, it is not possible to decouple several interactions simultaneously with different $\lambda$ spacing for each type. It is, of course, possible to decouple electrostatic and LJ interactions in a sequence, that can be defined via  \texttt{coul-lambdas} and \texttt{vdw-lambdas}, respectively, whereas the other is set to either zero (full interaction) or one (no interaction) for all intermediate states. %However, two separate simulation sets, as well as two \texttt{deltag-estimate} values. 

\section{Example and test cases}

When VI is switched off, all interactions are calculated as in Eqs.~(\ref{eq:ham_sim}), (\ref{eq:force_sim}) and (\ref{eq:dhdl_sim}). To test that VI collects all contributions correctly, for the following options in the mdp file,

\begin{lstlisting}[backgroundcolor = \color{light-gray}]
variational-morphing   = 1 
linear-test            = 1
\end{lstlisting}

Gromacs-VI calculates the intermediate Hamiltonian based on, 

\begin{align}
\begin{split}
H^\lambda_{VI}(\x, \lambda) \: =\; (1-\lambda) &\underbrace{\sum_{\substack{interaction \\type \;\; k  }} ... \sum_{\substack{particles \\i, j}} H^ k_A(\x_{i,j}, \lambda)}_{H^\lambda_A(\x, \lambda)} \\
+ \lambda &\underbrace{\sum_{\substack{interaction \\type \;\; k  }} ... \sum_{\substack{particles \\i, j}} H^ k_B(\x_{i,j}, \lambda)}_{H^\lambda_B(\x, \lambda)}  \;,
\label{eq:separate_sum}
\end{split}
\end{align}

and likewise, for the forces and $\lambda$ derivatives. Setting the seed to a fixed value such as, 
\begin{lstlisting}[backgroundcolor = \color{light-gray}]
ld-seed   =   1
\end{lstlisting}
it can be validated that all energies required for the free energy calculation that are stored in the \texttt{dhdl.xvg} file match between the implementation of the VI and the conventional sequence. \\ 

\begin{figure}
	\centering
	\includegraphics[width=0.3\linewidth]{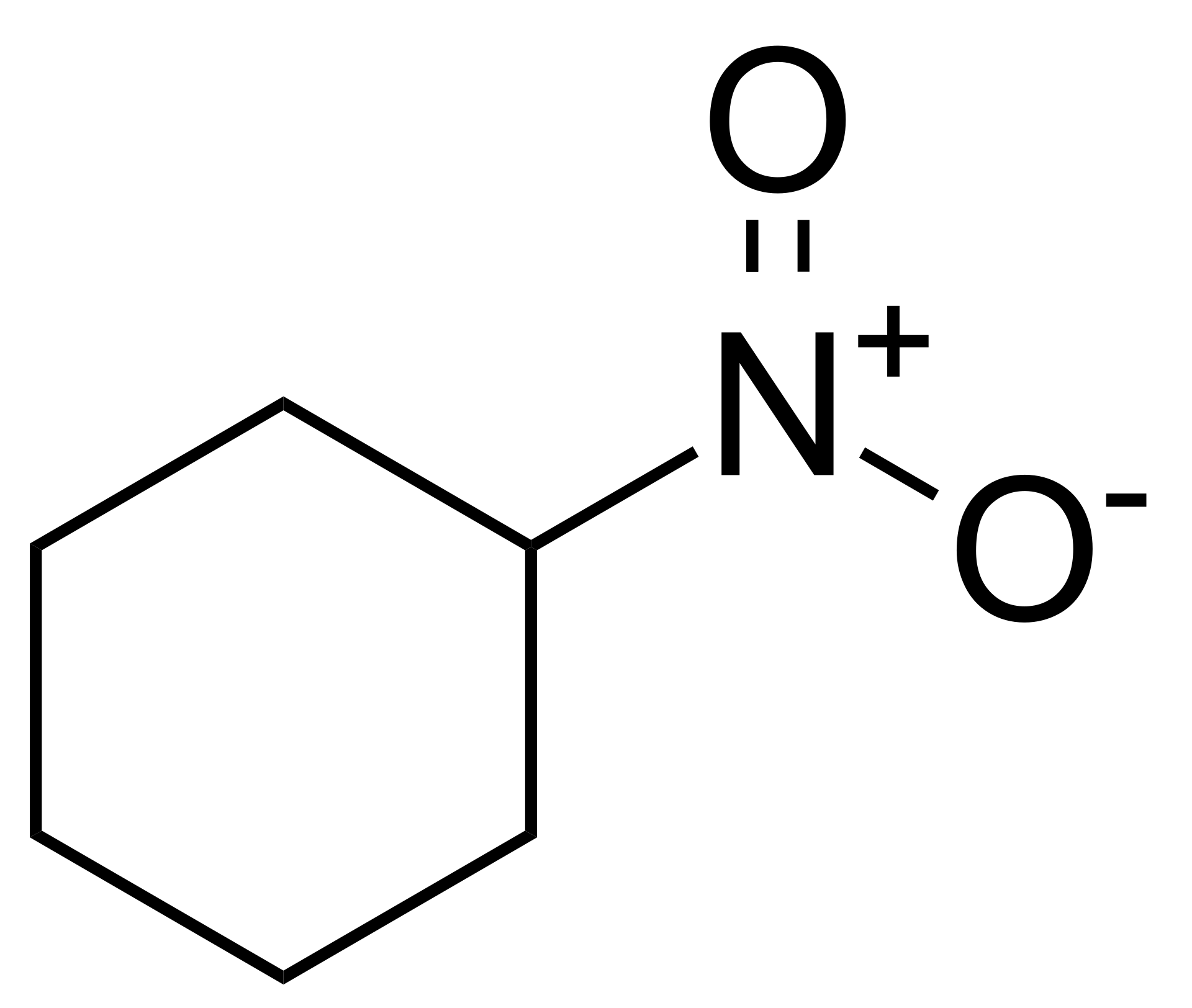}
	\caption{Structure of nitrocyclohexane, which is used as an example case.}
	\label{fig:nitrocyclohexane_structure}
\end{figure}

\subsection{Equilibrium States}

As an example case, the solvation free energy of nitrocyclohexane in water was calculated (structure shown in Fig.~\ref{fig:nitrocyclohexane_structure}). The topologies of the solvation toolkit package \cite{Bannan2016} created with the Generalized AMBER Force Field \cite{Wang2004} were used. Upon energy minimization, 2~ns NVT (constant volume and temperature) and 4~ns NPT (constant pressure and temperature) equilibration were conducted, followed by 100~ns production runs. \\

To asses whether the VI implementation yields accurate results consistent with the ones from conventional intermediates, first, through extensive sampling with 101 states (i.e., $\lambda$ steps of 0.01), a reference value value of (9.85 $\pm$ 0.02)~kJ$\slash$mol was obtained. It can be divided into (10.46 $\pm$ 0.01)~kJ$\slash$mol electrostatic, and (-0.61 $\pm$ 0.02)~kJ$\slash$mol LJ contributions. Next, a set of simulations with 5 states, i.e., $\lambda$ steps of 0.25, were conducted. \\

\begin{figure*}
	\subfloat{		\includegraphics[width=0.49\textwidth]{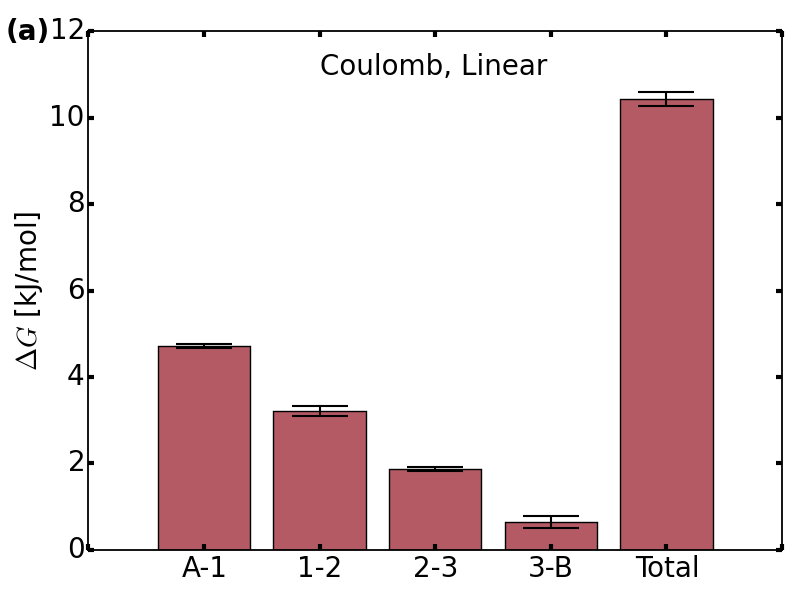}\label{fig:steps_linear_coulomb}} 
	\subfloat{		\includegraphics[width=0.49\textwidth]{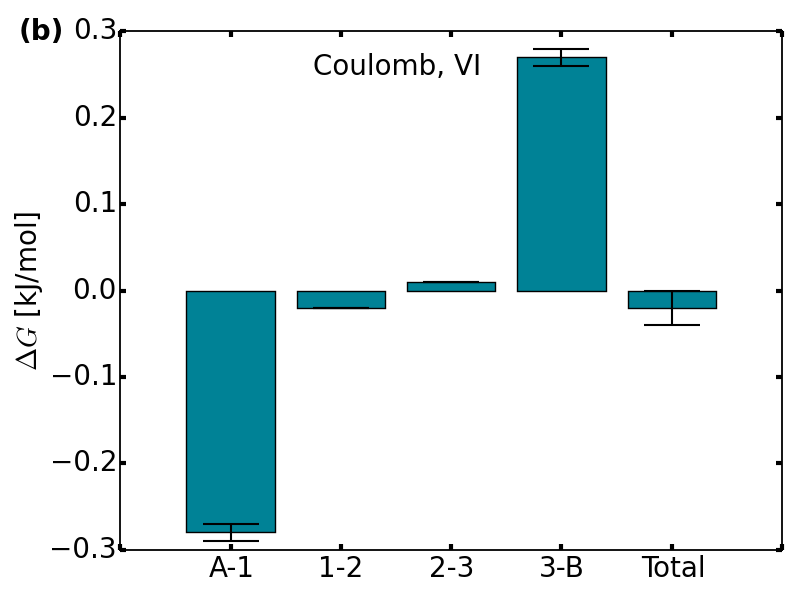}\label{fig:steps_vm_coulomb}}\\
	\subfloat{		\includegraphics[width=0.49\textwidth]{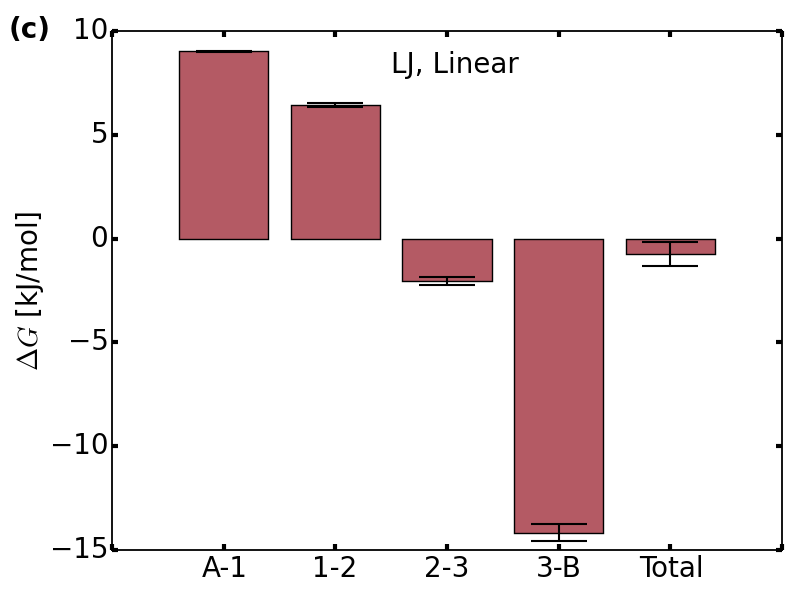}\label{fig:steps_linear_vdw}} 
	\subfloat{		\includegraphics[width=0.49\textwidth]{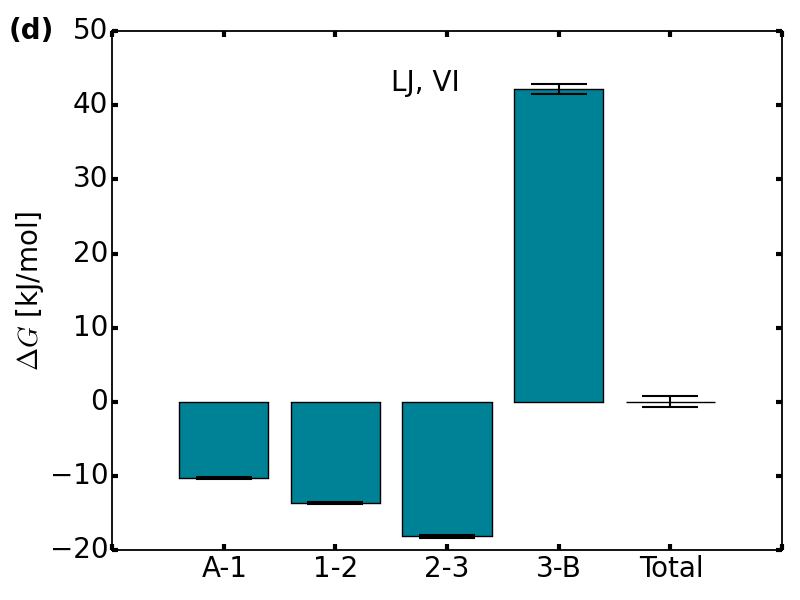}\label{fig:steps_vm_vdw}}
	\caption{Free energy differences along intermediate states between $A$ (coupled state) and $B$ (decoupled state). The bars show the differences between the states denoted below. The conventional linear interpolation method, panels (a) and (c), is shown in red, whereas VI is shown in  blue (panels (b) and (d)). Coulomb interactions were decoupled first (with LJ interactions still turned on), LJ interactions second (Coulomb interactions switched off).}\label{fig:steps}
\end{figure*}

The distribution of the free energy estimates between the different states is shown for Coulomb and LJ interactions in Fig.~(\ref{fig:steps}) and differs considerably between the two methods. The bars denote the free energy difference between the states denoted at the bottom. Again, $A$ represents the coupled, and $B$ the decoupled state. The plots shown for VI were created based on the runs where $C$ was set to the respective reference value, and, as such, sum up to about zero. When decoupling Coulomb interactions with a conventional linear interpolation method, shown in panel (a), the largest differences between the states occur in the first steps and gradually decreases. For VI (b), the free energy path along the intermediates has be become very small (note the differing unis on the axis). In contrast, for LJ interactions, the differences for VI (d) become larger than for the linear interpolation (c). The reason is, most likely, that the differences in the contributions from the attractive and the repulsive part of the LJ potential don't cancel for all intermediates.\\

\begin{figure}
	\centering
	\includegraphics[width=4in]{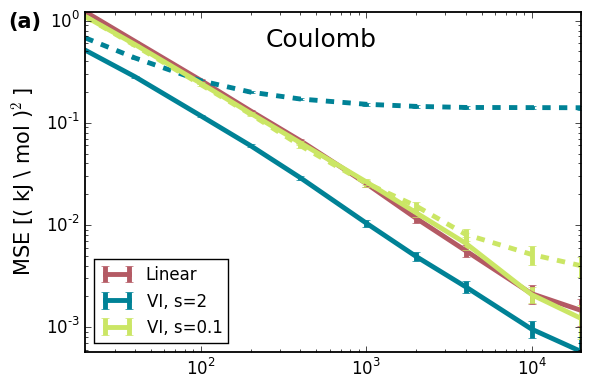}
	%\caption{MSEs of decoupling Coulomb interactions. }
	\includegraphics[width=4in]{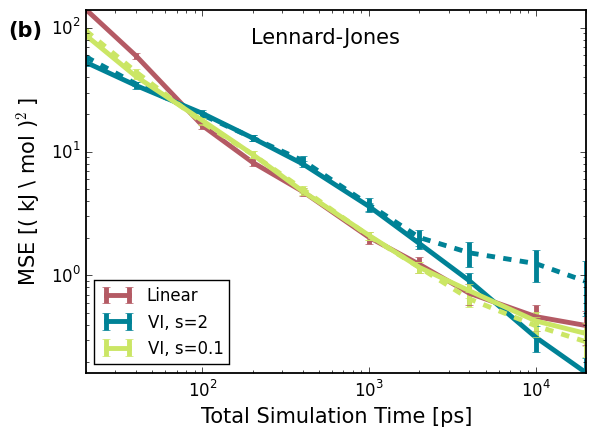}
	\caption{MSEs as a function of simulation time for decoupling (a) Coulomb and (b) Lennard-Jones interactions. The red line indicates the use of the conventional linear interpolation method, the blue and green line the VI approach, Eq.~\ref{eq:vi_all}, using two different $s$ values. The solid line indicate the MSEs that were obtained by using an exact initial guess, whereas a guess of 1~kJ$\slash$mol is indicated by the dashed lines. }
	\label{fig:mses}
\end{figure}

To compare the accuracy of both methods, Fig.~\ref{fig:mses} shows the MSEs with total simulation time, distributed equally over all five states. The MSEs were obtained by dividing the trajectories of the production runs into smaller ones, and comparing the resulting free energy difference to the reference value. For VI, two different smoothing values were considered (blue and green lines), as well as an exact initial estimate (solid line) and one that is 1~kJ$\slash$mol too low (dashed lines). \\

For electrostatic interactions, the MSEs in Fig.~\ref{fig:mses}(a) are significantly better for VI with $s=2$ and an estimate close to the exact one than the MSE obtained with linear intermediates, thereby validating the result of Ref.~ \citenum{Reinhardt2020}. However, in this case the MSEs are quite sensitive to the initial guess. For Lennard-Jones interactions, Fig.~\ref{fig:mses}(b), VI and linear intermediates yield similar MSEs, but the VI estimates are less sensitive to the initial guess. In both cases, the MSEs corresponding to VI with a smoothing factor of 0.1 are close to the linear ones and insensitive to the initial guess for most of the trajectory lengths in Fig.~\ref{fig:mses}. As such, it is advantageous to start the iteration process with a smaller smoothing factor that is gradually increased with an improved estimate for $C$.\\

\FloatBarrier
\section{Summary}
\label{sec:summary}

We have implemented the VI sequence of states into the GROMACS MD software package. For Coulomb interactions, our implementations yields significantly smaller MSEs and, in this sense, higher accuracy as compared to linearly interpolated intermediates. This results requires a sufficiently accurate initial estimate, which for the test cases presented here requires only a few percent of the overall simulation time. Furthermore, using the $\lambda$ dependence of the end states added to VI, for LJ interactions, similar MSEs as for conventional soft-core approaches are achieved. Given the many stepwise improvements that eventually led to the accuracy of current soft-core protocols, the fact the VI approach achieves similar accuracy already in the first attempt suggests that future refinements, e.g., of the lambda dependency on the end states, will push the accuracy even further.

%In contrast to the long list of improvements that led to the state of the art approaches and soft core potentials, the VI approach achieved similar accuracy off the cuff, without any further refinement. We therefore expect that future refinements of the lambda dependency of the end states will yield further substantial accuracy enhancements. 

\FloatBarrier
\section{Code and Data Availability}

The source code is available at 
%\color{blue}
 \url{https://www.mpibpc.mpg.de/gromacs-vi-extension} or \url{https://gitlab.gwdg.de/martin.reinhardt/gromacs-vi-extension}. Documentation, topologies and input parameter files of the above test cases are also available on the website and the repository. In the gitlab repository, all changes with respect to the official underlying GROMACS code can be retraced.  \\
 
  As installation is identical to that of GROMACS 2020, refer to \url{http://manual.gromacs.org/documentation/2020/install-guide/index.html} for detailed instructions.

\section{Acknowledgments}

The authors thank Dr. Carsten Kutzner for help, discussions and advice on GROMACS code development. 


%% References

\FloatBarrier
\bibliographystyle{elsarticle-num}
\bibliography{g_vi}




%% \ must have the following form:


% \end{thebibliography}

\end{document}